\PassOptionsToPackage{table,xcdraw}{xcolor}
\documentclass[sigconf]{acmart}

\AtBeginDocument{%
  }

\copyrightyear{2024}
\acmYear{2024}
\setcopyright{rightsretained}
\acmConference[CODS-COMAD Dec '24]{8th International Conference on Data Science and Management of Data (12th ACM IKDD CODS and 30th COMAD)}{December 18--21, 2024}{Jodhpur, India}
\acmBooktitle{8th International Conference on Data Science and Management of Data (12th ACM IKDD CODS and 30th COMAD) (CODS-COMAD Dec '24), December 18--21, 2024, Jodhpur, India}
\acmPrice{}
\acmDOI{10.1145/3703323.3704274}
\acmISBN{979-8-4007-1124-4/24/12}



\usepackage{booktabs,lipsum}
\usepackage{multirow}
\usepackage{balance}

\begin{document}

\title{Genicious: Contextual Few-shot Prompting for Insights Discovery}


 \author{Vineet Kumar}
 \orcid{0000-0001-8364-3981}
\affiliation{%
 \institution{PayPal}
 \city{Bengaluru}
 \country{India}
 }
 \email{vkumar32@paypal.com}

 \author{Ronald Tony}
 \orcid{0009-0005-5772-1065}
\affiliation{%
 \institution{PayPal}
 \city{Bengaluru}
 \country{India}
 }
  \email{rtony@paypal.com}

 \author{Darshita Rathore}
 \orcid{0000-0003-0430-1129}
\affiliation{%
 \institution{PayPal}
 \city{Bengaluru}
 \country{India}
 }
  \email{drathore@paypal.com}

 \author{Vipasha Rana}
 \orcid{0009-0001-5447-6587}
\affiliation{%
 \institution{PayPal}
 \city{Bengaluru}
 \country{India}
 }
   \email{viprana@paypal.com}

 \author{Bhuvanesh Mandora}
 \orcid{0009-0005-1776-1724}
\affiliation{%
 \institution{PayPal}
 \city{Bengaluru}
 \country{India}
 }
  \email{bmandora@paypal.com}

 \author{Kanishka}
 \orcid{0009-0003-3208-945X}
\affiliation{%
 \institution{PayPal}
 \city{Bengaluru}
 \country{India}
 }
  \email{kdhar@paypal.com}

 \author{Chetna Bansal}
 \orcid{0009-0007-2649-3770}
\affiliation{%
 \institution{PayPal}
 \city{Bengaluru}
 \country{India}
 }
  \email{cbansal@paypal.com}

 \author{Anindya Moitra}
 \orcid{0009-0007-8165-0003}
\affiliation{%
 \institution{PayPal}
 \city{Bengaluru}
 \country{India}
 }
   \email{amoitra@paypal.com}

\renewcommand{\shortauthors}{Kumar et al.}

\begin{abstract}
Data and insights discovery is critical for decision-making in modern organizations. We present Genicious, an LLM-aided interface that enables users to interact with tabular datasets and ask complex queries in natural language. By benchmarking various prompting strategies and language models, we have developed an end-to-end tool that leverages contextual few-shot prompting, achieving superior performance in terms of latency, accuracy, and scalability. Genicious empowers stakeholders to explore, analyze and visualize their datasets efficiently while ensuring data security through role-based access control and a Text-to-SQL approach.
\end{abstract}




\maketitle

\section{Introduction}

In today's digital age, data has become an essential resource. However, the sheer volume and growing complexity of data present significant challenges in querying and exploration, even for seasoned professionals. While leaders and senior leadership require quick access to insights about key performance indicators (KPIs), traditional methods of data presentation, such as dashboards, often fall short of providing comprehensive, flexible analysis. These pre-designed templates, though useful, constrain users to a limited set of predefined queries and visualizations. This rigidity becomes a significant barrier when decision-makers need to explore data dynamically, ask ad-hoc questions, or investigate nuanced aspects of their business that don't conform to existing templates. 

\textsc{Text-to-SQL} interfaces address this limitation by allowing users to pose questions in natural language, breaking free from the constraints of pre-built dashboards. This capability empowers leaders and team members alike to query data spontaneously, uncovering deeper insights and responding to emerging trends or challenges in \textit{real-time}. By bridging the gap between complex database structures and intuitive human inquiry, \textsc{Text-to-SQL} systems not only enhance data accessibility but also foster a culture of data-driven decision-making across all levels of an organization, truly unlocking the power of data analytics.

However, \textsc{Text-to-SQL} is a longstanding problem in the domain of Natural Language Processing. Formally, given a natural language query $(Q)$ and its corresponding database schema $(S)$, which presents data and attribute type information, a \textsc{Text-to-SQL} system aims to generate a valid and executable SQL query $(y)$, which when executed on the database; will return results that match the user’s original intent.

One of the primary reasons for framing insights discovery as a \textsc{Text-to-SQL} process is the need to maintain data confidentiality. In many cases, the datasets being queried contain sensitive or proprietary information that cannot be shared with external language models (LLMs). By utilizing a \textsc{Text-to-SQL} approach, the actual data remains secure and undisclosed, as only the schema or metadata of the database is exposed.

\textsc{Text-to-SQL} acts as a middle ground where the LLM only needs access to the database structure—not the data itself—to generate SQL queries. These SQL queries can then be executed internally to retrieve insights, ensuring that the raw data never leaves the secure environment. This approach allows users to ask questions and gain insights about specific datasets while maintaining strict data protection and compliance with privacy policies.

Recently, the evolution of large language models (LLMs) has reshaped the landscape. Models such as GPT and Llama show exceptional abilities and achieve very good results in various tasks with minimal fine-tuning. These models are particularly effective in few-shot and zero-shot learning contexts, enabling them to tackle complex tasks with little or no additional training data. This advancement has expanded the potential for applying LLMs in more sophisticated and diverse applications for tabular data.

Though LLMs have demonstrated impressive capabilities, they continue to struggle with limitations, particularly in handling domain-specific or knowledge-intensive tasks. These challenges often manifest as `hallucinations' where the models generate incorrect or misleading information. To address these issues, Retrieval-Augmented Generation (RAG) has been developed as a solution that strengthens LLMs by retrieving relevant information from external knowledge sources through semantic similarity matching. By incorporating this external data, RAG significantly reduces the occurrence of factual inaccuracies in generated content. 

To address the aforementioned challenges and cater to a diverse user base - from SQL novices to experts across various business functions - we have developed \textbf{Genicious}\footnote{Demo is available \href{https://drive.google.com/drive/folders/1EIV0gbevofyIgn4oqmLt2caQJRIkMOhc}{here}.}, an innovative Gen AI-powered system for data querying and insights generation. Genicious leverages the text-generative capabilities of LLMs through a novel approach to bridge the gap between natural language communication and complex database interactions. This system is designed to democratize data access, enabling users of all skill levels to extract valuable insights from relational databases through intuitive, conversational interfaces.

Here are the key contributions of our paper ---
\begin{itemize}
    \item We conducted a comprehensive benchmarking of several widely used open-source and proprietary LLMs for SQL generation, evaluating their performance on both public and our domain-specific datasets. 
    \item We explored various prompting strategies aimed at enhancing the efficiency and accuracy of SQL code generation.
    \item Drawing on experimental evidence, we implemented a novel Contextual Few-shot Prompting approach by integrating RAG to dynamically adapt few-shot examples based on user queries, thereby enhancing SQL generation performance.

\end{itemize}

The rest of this paper is organized as follows -- In Section \ref{Related Works}, we briefly discuss related works. Section \ref{expt} talks about various experiments we conducted and the rationale for the design choices we made. Section \ref{system} presents the system descriptions. Finally, Section \ref{conclusion} concludes the paper with a few possible directions for future improvement \& research.

\section{Related Works}\label{Related Works}
This field has witnessed significant advancements in the last decade, evolving from rule-based complex feature engineering systems to very recent large language models (LLMs). Hong et al. \cite{hong2024next} and Shi et al. \cite{shi2024survey} provide an excellent survey of methods in this direction.

 Early \textsc{Text-to-SQL} systems were predominantly rule-based \cite{li2014constructing}\cite{mahmud2015rule} and often involved extensive feature engineering. While these systems demonstrated great performance in certain domains (for which they were \textit{trained}) they lacked the generalization capability.

With advancements in Deep Learning, Seq2Seq models have shown promising results in the domain. The appeal of these models \cite{sutskever2014sequence}\cite{xu2017sqlnet} lies in their ability to transform natural language inputs into corresponding SQL outputs directly. This end-to-end approach significantly streamlines the process, eliminating the need for extensive feature engineering and the creation of complex rule-based systems. 

Remarkable text generation capabilities of LLMs\cite{brown2020language} have shown very promising results in this domain \cite{dong2023c3}\cite{pourreza2024din}. Broadly there are two main approaches to use LLMs for SQL generation --  prompt engineering \cite{chang2023prompt} and fine-tuning \cite{gao2023text}. To rigorously evaluate the accuracy of these systems, researchers have developed several benchmark datasets. Among these, Spider \cite{yu2018spider} described as \emph{a large-scale complex and cross-domain semantic parsing and \textsc{Text-to-SQL} dataset} -- has become a standard for measuring progress in the field.
Currently, leading approaches on the Spider leaderboard are C3\cite{dong2023c3}, DIN-SQL\cite{pourreza2024din},
and DAIL-SQL\cite{gao2023text}.

\section{Experiments \& Design Choices}\label{expt}

While developing our Natural Language-based Insights Discovery tool, we faced three crucial decisions that significantly impacted the system's performance and usability. These key aspects were -- 
\begin{itemize}
    \item Selection of an appropriate Large Language Model (LLM)
    \item Design of a scalable \& accurate prompting strategy
    \item Choice of evaluation criteria for model-generated outputs
\end{itemize}
Each of these decisions required careful consideration of various options, balancing factors such as performance, cost-effectiveness, and scalability. In this section, we will elaborate on the alternatives we explored for each aspect, the choices we ultimately made, and the rationale behind our decisions.

\subsection{Problem Formulation} Given a natural language query ${Q}$ and its corresponding database schema ${S}$, (which presents data type and attribute type information), a \textsc{Text-to-SQL} system (that utilizes LLMs) aims to generate a valid and executable SQL query $y$, when query and schema are fed into LLM in a prompt template $\mathcal{T}(Q,S)$. Formally, the auto-regressive generation process of SQL query $y$ can be formulated as follows:
\[ \Pr\left(y|\mathcal{T}(Q,S)\right)=\prod_{i=1}^{|y|}\Pr\left(y_i|\mathcal{T}(Q,S),y_{<i}\right)\]

\subsection{Choice of LLMs}\label{l}
We chose both popular open-source as well as proprietary LLMs for our experiments. For open-source LLMs, a thorough review by the legal team is done to ensure compliance, and upon approval, the LLMs are added to the repository and made available for use. Due to legal constraints, only LLMs that have been approved through this process are included. Specifically, we selected versions of Llama \cite{touvron2023llama}, GPT \cite{achiam2023gpt}, and Mistral \cite{jiang2023mistral} for our analysis. More details about models and their specifications of parameter size and model type (whether they are foundation model or assistant model) is described in Table \ref{tab:llm}. All open-source models are full-precision variants. 

\begin{table}
\begin{tabular}{@{}lll@{}}
\toprule
Model         & Size  & Type     \\ \midrule
Llama 2       & 13 B   & Chat     \\
Mistral       & 7 B   & Instruct \\
Llama 3       & 8 B   & Instruct \\
Llama 3.1     & 8 B   & Instruct \\
GPT 3.5 Turbo & 175 B & Chat     \\
GPT 4         & 1.8 T & Chat     \\ \bottomrule
\end{tabular}
 \caption{Candidate LLMs for evaluation}
  \label{tab:llm}
  \normalsize
  \vspace{-0.4in}
\end{table}

\subsection{Prompt Engineering}\label{pro}
Prompt engineering constitutes a strategic combination of information such as instructions, questions and the addition of table information, aimed at enhancing the performance of LLMs to generate expected output. We explored advanced prompt engineering techniques that enable us to tackle complex tasks while enhancing the LLM's reliability using basic instruction, supplementary knowledge and ensembling a few techniques as follows: \\
Reasoning:
\begin{itemize}
    \item Prompt Chaining: Involves linking multiple prompts together, where the output of one prompt becomes the input for the next, to guide a model through complex tasks.
    \item Self-Consistency: Involves generating multiple responses to the same prompt and selecting the most coherent answer to improve reliability.
    \item Cross-Consistency: Involves comparing responses generated from different models to identify and select the most consistent or accurate answer.
\end{itemize}

Implementation of LLM-based \textsc{Text-to-SQL} systems often relies on In-context learning \cite{brown2020language}. ICL offers Language models to learn tasks given the instructions and a few examples in the form of demonstrations.

\begin{itemize}
    \item Zero-shot Prompting: Model tries to generate SQL queries without any additional examples.
    \item Few-shot Prompting: Model is given a few examples to serve as a foundation for guiding it to generate responses for better performance.
\end{itemize}

However, LLMs often encounter limitations in generating accurate output in cases where access to domain-specific data and specific private information is required. Unlike publicly accessible sources, proprietary data is often stored in relational SQL databases,  PDFs, Confluence pages, slide decks, etc. Fine-tuning LLMs, which requires customizing the model using new datasets for a particular task, comes with challenges such as high training costs, powerful system configuration, sending data to the model, and limited transparency in the model's decision-making. Hence, there is a need for an approach which relies on learning through contextual information, additional relevant data and enhancement through Retrieval-Augmented Generation (RAG) \cite{fan2024survey} to enrich the context leading to more accurate and contextually appropriate outputs.

Using RAG we can make the demonstrations of few-shot prompting more contextual. Based on user query we can retrieve the most relevant and contextual examples to be fed as a system demonstration for SQL generation.

\begin{figure}[h]
    \centering
    \includegraphics[width=0.45\textwidth]{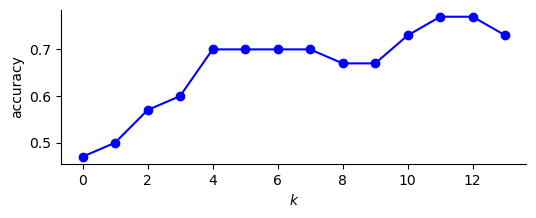}

    \caption{Number of Examples$(k)$ in Context vs. Accuracy}
    \label{fig:acc_fk}
\end{figure}

For few-shot learning, to determine the optimal number of examples $(k)$ to include in the prompt, we conducted a brief analysis by evaluating accuracy across different values of $(k)$. Our findings indicated that beyond $k=4$, the improvement in accuracy is not prominent. Additionally, increasing the number of tokens can incur higher computational costs for some models. Therefore, we selected $k=4$ as the optimal value, as illustrated in Figure \ref{fig:acc_fk}.

\subsection{Evaluation of Different LLMs \& Prompting Strategies}
To evaluate the performance of the various LLMs and prompting strategies described in Section \ref{l} and \ref{pro}, we conducted a benchmarking experiment. For this experiment, we utilized a very simple prompt template--  
\begin{verbatim}
    <prompt template>
    {System Instructions}
    {Database Schema}
    {Optional Demonstrations}
    </prompt template>
\end{verbatim}

To conduct our experiments we used the Spider \cite{yu2018spider} dataset and evaluated the accuracy of LLM generated model using the following two criteria (both of these methods are from the Spider evaluation framework)
\begin{enumerate}
    \item Exact Match of generated SQL: We decompose each SQL into several clauses, and conduct a set comparison in each SQL clause
    \item Execution Accuracy: We compare the values of the predicted SQL queries with those of the ground truth SQL query
\end{enumerate}

\begin{figure}[h!]
    \centering
      \includegraphics[width=1.\linewidth]{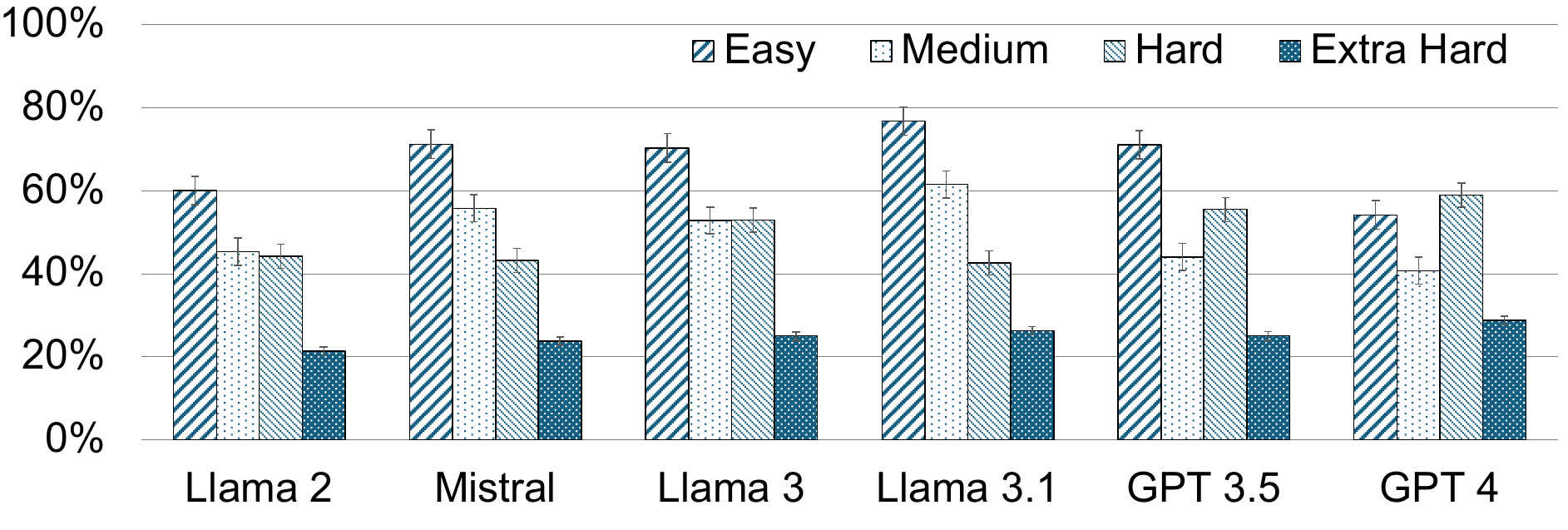}
    \caption{Performance of LLMs wrt. difficulty level on Spider \cite{yu2018spider} test set.}
    \label{fig:diff}
\end{figure}

We found that both of these metrics are highly correlated. Therefore, we are only reporting Execution Accuracy in Table \ref{tab:commands}. Few-Shot Prompting consistently outperforms Zero-Shot, and Contextual Few-Shot outperforms both methods. We also experimented with self-consistency; however, due to the increased latency and minimal accuracy gains from querying the model multiple times, we chose Contextual Few-Shot Prompting as our preferred strategy. Figure \ref{fig:diff} illustrates the comparative performance of LLMs with respect to the difficulty of questions. GPT is performing better for \textit{hard} questions.

Among all competing models, Llama 3.1 proved to be the most efficient and accurate in our experiments. However, when tested on our domain-specific dataset, GPT-3.5 Turbo emerged as the top-performing model.  Hence we chose GPT-3.5 Turbo for developing Genicious.

\begin{table}[h!]
\resizebox{\columnwidth}{!}{%
\begin{tabular}{@{}lcccccc@{}}
\toprule
Strategy  & Llama 2 & Mistral & Llama 3 & Llama 3.1                                & GPT 3.5 T                                & GPT 4   \\ \midrule
ZS        & 5.0\%  & 36.0\% & 27.0\% & \cellcolor[HTML]{CAEDFB}\textbf{41.0\%} & 37.0\%                                  & 28.0\% \\
FS        & 51.0\% & 51.0\% & 52.0\% & 53.0\%                                  & \cellcolor[HTML]{CAEDFB}\textbf{53.0\%} & 52.0\% \\
CFS       & 68.0\% & 70.0\% & 69.0\% & \cellcolor[HTML]{CAEDFB}\textbf{71.0\%} & 65.0\%                                  & 55.0\% \\
SC        & 26.0\% & 40.0\% & 51.0\% & \cellcolor[HTML]{CAEDFB}\textbf{53.0\%} & 53.0\%                                  & 47.0\% \\
CFS w/ SC & 70.3\% & 66.7\% & 70.0\% & \cellcolor[HTML]{CAEDFB}\textbf{70.3\%} & 60.0\%                                  & 56.7\% \\ \bottomrule
\end{tabular}
}
  \caption{Evaluation of different LLMs \& Prompting Strategies on Spider \cite{yu2018spider} test set. \normalfont ZS: {Zero Shot}, FS: Few Shot$(k=4)$, CFS: Contextual Few Shot $(k=4)$, SC: Self Consistency $(n=5)$, CFS w/ SC: Contextual Few Shot w/ Self Consistency $(k=4,n=5)$.}\vspace{-0.35in}
  \label{tab:commands}
\end{table}

\section{System Description}\label{system}
\begin{figure}[h]
  \centering
  \includegraphics[width=\linewidth]{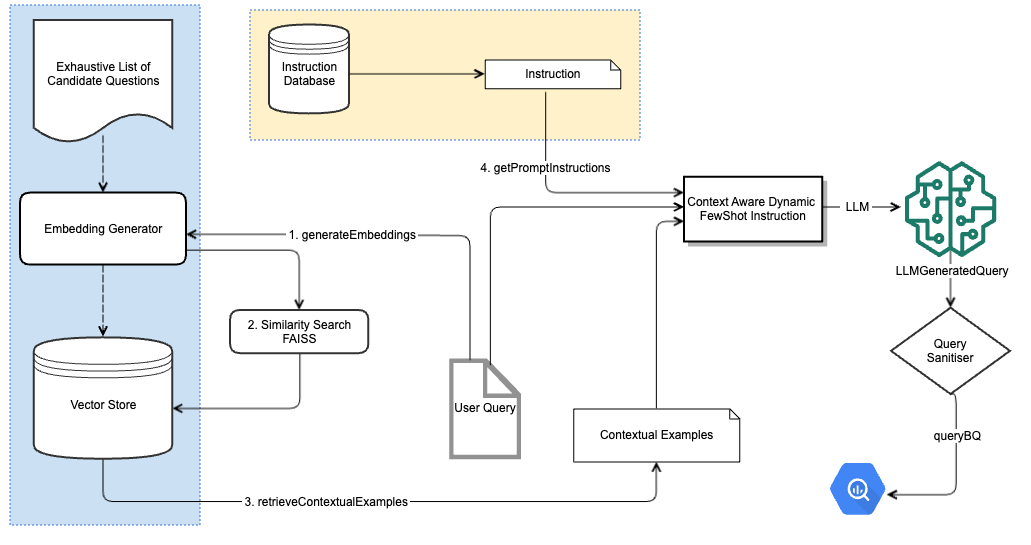}
  \caption{Tool Architecture}
  \label{arch}
  \vspace{-0.1in}
\end{figure}

Genicious is a Java Spring Boot–based application with a React.js user interface, allowing users to interact and communicate through REST APIs for querying various datasets by leveraging advanced natural language processing techniques. The architecture of the Genicious tool is shown in Figure \ref{arch}. There are two broad components in the tool -- one is an offline phase (shaded in blue) and the other is a query phase.

\paragraph{Offline Phase: Onboarding and Vector store Preparation}

During the onboarding of a new dataset, a list of candidate questions and their corresponding SQLs are gathered. These questions should exhaustively cover various \textit{themes} of the domain. For example, when using the tool to answer questions involving year-over-year comparisons, a similar question and its corresponding SQL query can be included in the candidate question pool. Once we have the question pool ready we use an embedding model to convert these questions into numerical embeddings. We are using \texttt{text-embedding-ada-002} model. These embeddings, which are numerical representations capturing the semantic essence of the questions, are stored in a vector database (Milvus), facilitating efficient retrieval during the query phase. 

 As new datasets/projects are introduced, the system is incrementally updated by adding corresponding embeddings of the question pool into the vector store. This dynamic approach ensures that the vector store remains current and relevant, continuously improving the accuracy and efficiency of the similarity search and query generation processes. 

By integrating these advanced techniques and maintaining an adaptive vector store, our system provides a robust and scalable solution for querying diverse projects and generating precise responses.

\paragraph{Query Phase: Input Processing and Query Generation}

Now, when a user  submits a query $Q$, the system performs the following steps:

\begin{enumerate}

\item \textbf{Embedding Generation:} The user's input question $(Q)$ is converted into a numerical embedding $(e)$ using the text embedding model \texttt{text-embedding-ada-002}. This embedding is then used to retrieve SQL examples that are relevant as part of the Retrieval Augmented Generation (RAG) framework.

\item \textbf{Similarity Search:} Using FAISS (Facebook AI Similarity Search) \cite{johnson2019billion}, we perform a similarity search with the embedding generated in the previous step $(e)$ against pre-stored question embeddings in the vector database. FAISS efficiently ranks the questions based on their \textit{proximity} to the user's question.

\item \textbf{Contextual Examples Retrieval:} This step retrieves and presents the most relevant textual questions from the question pool, mapped back from their numerical embeddings. These retrieved questions are used to generate contextualized examples where the few-shot demonstrations are dynamic (adapting to the input question) and relevant.

\item \textbf{Prompt Building:} We maintain an instruction database that stores prompt templates for each project/dataset. For every user query, an appropriate instruction prompt is fetched from this database and appended with the original user query and retrieved contextual few-shot examples. This context-aware, dynamic few-shot instruction is then sent to the LLMs.

\item \textbf{LLM Query Generation:} The LLM formulates a BigQuery (BQ) query tailored to effectively address the user's question.

\item \textbf{SQL Sanitization:} The LLM-generated SQL undergoes a sanitization step, which involves scanning the SQL for any harmful keywords (e.g., \texttt{DROP, ALTER, UPDATE}, etc.).

\item \textbf{Response Delivery:} If the SQL is deemed safe, the query is executed on the relational database, and the results are returned via REST API calls, ensuring a seamless and responsive interaction. Finally, the results are presented to the user using tables and visualizations.

\end{enumerate}

\section{Adoption and Future Steps}\label{conclusion}

The P95 latency for a single query response is around 6 seconds. The tool efficiently handles simple to intermediate queries on flat tables, retrieves the latest metrics, performs basic calculations, and investigates trends. We are actively working on adding new features in future version such as making the bot truly conversational with scoped interactions, and utilizing an agentic framework while ensuring robust data protection, among others.


\balance
  \bibliographystyle{ACM-Reference-Format}
  \bibliography{auths}
\end{document}